# Quantifying the Complexity of Materials with Assembly Theory

Keith Y Patarroyo[1†], Abhishek Sharma[1†], Ian Seet[1], Ignas Pakamorė[1], Sara I. Walker[2,3] and Leroy Cronin[1,3]*

[1]School of Chemistry, The University of Glasgow, University Avenue, Glasgow G12 8QQ, UK

[2]Beyond Center for Fundamental Concepts in Science, Arizona State University, Tempe AZ, USA

[3]Santa Fe Institute, Santa Fe NM, USA

*Corresponding author email: Lee.Cronin@glasgow.ac.uk [†]Equal contribution.

**Abstract:** Quantifying the evolution and complexity of materials is of importance in many areas of science and engineering, where a central open challenge is developing experimental complexity measurements to distinguish random structures from evolved or engineered materials. Assembly Theory (AT) was developed to measure complexity produced by selection, evolution and technology. Here, we extend the fundamentals of AT to quantify complexity in inorganic molecules and solid-state periodic objects such as crystals, minerals and microprocessors, showing how the framework of AT can be used to distinguish naturally formed materials from evolved and engineered ones by quantifying the amount of assembly using the assembly equation defined by AT. We show how tracking the Assembly of repeated structures within a material allows us formalizing the complexity of materials in a manner accessible to measurement. We confirm the physical relevance of our formal approach, by applying it to phase transformations in crystals using the HCP to FCC transformation as a model system. To explore this approach, we introduce random stacking faults in closed-packed systems simplified to one-dimensional strings and demonstrate how Assembly can track the phase transformation. We then compare the Assembly of closed-packed structures with random or engineered faults, demonstrating its utility in distinguishing engineered materials from randomly structured ones. Our results have implications for the study of pre-genetic minerals at the origin of life, optimization of material design in the trade-off between complexity and function, and new approaches to explore material technosignatures which can be unambiguously identified as products of engineered design.

**Introduction**

The process of self-organisation directs the formation of complex structures,[1-2] where the dissipation of energy[3] leads to the organisation of matter across scales of space and time.[4,5] However, the emergence of components necessary for evolving systems[6] requires involves an increase in complexity at the molecular scale.[7] Discovering features of living systems and designing novel materials therefore relies on concrete notions of molecular complexity[8-9] that can be tested experimentally. Of particular interest for the study and design of new living and synthetic materials is the recognition of chemistry as the first scale at which combinatorial complexity can be manifested and observed. There exist a vast large number of possible molecular objects, which could be synthesized on Earth via the combination of local abundant elements using different bonding configurations.[10]

Inorganic materials, including minerals[11] and clays[12], are important catalytic systems that can help facilitate all manner of different types of chemical reactions. These systems are important because the structural aspects e.g. cavity size, presence of reactive sites, or defects have also been suggested to be a vital component for the emergence of life on Earth. This is because the particular configuration of certain abiotic materials has been suggested to have helped kick-started the process of evolution.[13] The presence of such complex configurations generated by happenstance has been hypothesised to help generate the initial frozen accident[14] that could have led to subsequently to genetic heredity and the evolution of the genetic code. This is because there is an open question about how such material complexity could be measured, calculated, and investigated, to test such hypotheses of the causation for genetic information that ultimately manifested at the origin of life from abiotic materials.

In general, in natural and synthetic inorganic molecules and materials, complexity emerges not only from the variety of components, such as their building blocks, but also in the selection of specific patterns and structures they might form[15], mediated by geochemical conditions, biological cells, or human engineers. As an example, at different temperatures, Titanium has two different crystal structures: hexagonal closed packed at room temperature and body-centred cubic at a high temperature, highlighting how materials with the same composition can yield different observable complexity.[16] It is often the case that human-engineered materials are made out of the same elements as naturally formed materials, leaving open the question of how to distinguish complexity associated to evolution and also intelligent engineering which are both related but different classes as one produces the other. Considering the periodicity within a material at hierarchical scales, starting from atomic arrangements to the architectural design of the entire sample, suggests it should be possible to differentiate a highly ordered material as a sign of technological production as distinct from objects formed from natural processes with disordered structure. This is relevant in the context of the origin of life because it suggests unstructured abiotic materials, e.g., the kind that may have preceded life, may have an upper bound on the order they can contain in their structure. If it can be shown this bound is surpassed by engineered materials, it also has implications for material technosignatures.

In general, inorganic systems are considered relatively simple due to their periodic and often symmetrical nature, yet subtle changes in atomic arrangements can yield structures that appear highly complex or random; indeed it is a hard problem to tell the difference between complexity and randomness.[17,18] Highly ordered inorganic molecules and metallic crystals are regarded as "less complex" as they can be represented by a unit cell and the number of periodic units within the sample. At the molecular scale, inorganic molecules are often composed of well-defined, discrete units such as small clusters or polyatomic ions.[19] These molecules, especially those containing high oxidation state metals, are described by a finite number of atoms connected by ionic or coordinate bonds.[17] Their perceived, or potential for, complexity arises from the bond type, coordination environments, and geometries that can be adopted by the constituent atoms, especially when transition metals or heavy elements are involved.[20] Extended inorganic materials, such as metallic crystals, metal-organic frameworks (MOFs)[21] or covalent organic frameworks (COFs)[22] no longer consist of discrete units but instead form continuous structures where atomic arrangements extend infinitely (or nearly so) in one, two, or three dimensions. These are characterised by long-range order and the periodicity of their atomic arrangement, as seen in materials like alloys, salts, oxides, or zeolites.[23] The complexity of such extended systems is not only determined by the local atomic environment but also by the larger repeating patterns and how they propagate in space across the material.

Quantifying the structural complexity of a crystalline material is an open challenge, and various approaches have been used to quantify the complexity of inorganic materials.[17,20,24] Information-theoretic approaches, utilising Shannon entropy[25], mutual information,[26] and various algorithmic complexity measures like Kolmogorov complexity[27] have been used to quantify the complexity of materials, but a method to directly measure complexity empirically is missing. The order or disorder within the structure of crystalline materials can be defined by informational entropy but this is not directly measurable. A highly ordered structure, such as a perfect crystal at absolute zero, corresponds to low entropy because there is little uncertainty in the arrangement of its components such as atomic positions. Conversely, a highly disordered system, like a glass or an amorphous solid exhibits high entropy. However, it is important to note that a material with high informational entropy may appear structurally complex, but that doesn't imply that it possesses unique mechanical, electrical, or chemical properties. Moreover, information entropy takes into account a static snapshot of a system without tracking the causal process that produced the system, while in reality, materials are dynamic and can undergo phase transitions, atomic rearrangements, or other processes that influence their properties over time.[28]

Assembly Theory (AT) was developed initially to distinguish biological samples from non-biological samples by utilising molecules as biosignatures[29]. It is the first experimentally measurable form of complexity, with a metrology of complexity rooted in spectroscopy.[30-31] Later, it was extended further to explain selection processes before the genome, with a goal of formalizing generalized evolutionary processes in an empirically tractable manner.[32] The foundational principles of AT can be applied to any physical system, once a precise definition of the object undergoing selection has been made. Within the context of AT, an object is defined by an entity that is finite, distinguishable, persists over time to be observed, and breakable such that the constraints required for its construction can be measured.[32] The complexity of an object, so defined, is quantified by the assembly index ($a_i$), which represents the minimum number of steps required to construct the object from its basic building blocks. Unlike calculation of information theoretic complexity measures, including those applied previously to crystalline materials,[33] assembly pathways are intended to capture physical constraints within a material and can be directly probed by measurement. The assembly pathway represents and quantifies minimum causal contingency associated with construction of object, which does not depend on how the material was synthesized or manufactured, but instead captures the captured constraints maintaining its stability as a configuration of matter. This is a particularly useful approach when mechanistic insights into the formation of an object are unknown (as is the case for classifying materials of unknown origin). When multiple objects are observed, the AT framework uses Joint Assembly Spaces (JAS),[34] which represent the minimal pathways required to simultaneously construct multiple objects. To quantify the degree of selection required to construct an ensemble of observed objects, AT utilises the assembly index and the copy number of objects, where selective constraints are quantified by the assembly ($A$),

$$A = \sum_{i=1}^{N} e^{a_i} \left(\frac{n_i - 1}{N_T}\right) \tag{1}$$

where $A$ is the assembly of the ensemble, $a_i$ is the assembly index of the $i^{th}$ object, $n_i$ is its copy number, $N$ is the total number of unique objects, $e$ is Euler's number and $N_T$ is the total number of objects in the ensemble.

The formation of crystalline materials from nucleation and growth processes leads to the formation of an extensive network of interconnected atoms arranged in a regular pattern, a structural insight affirmed through X-ray diffraction and electron microscopy experiments. Here we explore the application of assembly theory to objects that are crystalline materials. The observed symmetry in crystalline materials (objects) as an outcome of the periodic arrangement of atoms can be compressed into a single representative unit as a unit cell or by an irreducible unit. In principle, any observed crystalline object can be regarded as a combination of the construction of irreducible units and their spatially periodic arrangement to create the observed object. A human-engineered architecture such as a CMOS chip[35] and agglomerated arrangement of nanoparticles are both periodic at the atomic scale and extremely complex at the object scale, see Figure 1A and 1B. However, state of the art silicon chips have trillions of precisely drawn features created by ultra-lithography, achieved by the careful deposition and removal of various layers of materials to create the features with the correct electronic properties and geometric connectivity. These features are not only approaching atomic limits, but they need to be robust to withstand high temperatures associated with the very fast and highly controlled passage of electrons through these architectures. These features are used to both form and connect billions of transistors, the basic units used to build logic gates. These logic gates in turn form the basis of all the nano-electronic architectures that make up most of the central processing unit with some memory elements, again made from lithographically created architectures. In this regard, engineered objects designed for specific functions are highly structured with an intricate balance between complexity and periodicity, yielding high copy numbers for internal structure. Within a silicon chip, it is possible to observe vast numbers of units, some of which are present as billions of copies with high assembly index due to the number of steps required to produce the units, see Figure 1C. In some regards, one can think of silicon chips as being the most advanced mineral types currently known in the universe, and the engineering required to produce these architectures represents a vast amount of research and development since the first desktop transistor was built in 1947.[36] By contrast, unstructured objects formed from random processes lead to extremely high complexity without periodicity at higher order, see Figure 1.

Herein, we start by expanding the fundamentals of AT to periodic, solid-state objects such as crystals and inorganic materials. We extend the quantification of assembly index and pathways by hierarchical decomposition of an object into its unit cell (or supercells) and their periodic arrangement to develop the theory for crystalline materials. As an application, we analyse the complexity of unit cells of various observed crystal structures from the American Mineralogist Crystal Structure Database.[37] Considering the periodicity of the unit cells (or supercell) within the object, we introduce a concept of internal copy number and describe its scaling with object size. To demonstrate how AT captures causal contingency in observed structures, we explore the joint assembly space representing the phase transition within carbon allotropes from graphite to diamond. We further expand the concept of assembly index and internal copy number to objects with point defects by introducing supercells to compensate for local complexity due to the presence of defects. As a final example, we consider HCP to FCC phase transformation by introducing stacking faults. We demonstrate how complexity emerges scales or organization within a material due to the presence of random or engineered faults, which could suggest new strategies for manipulating material properties and classification of materials of unknown origin.

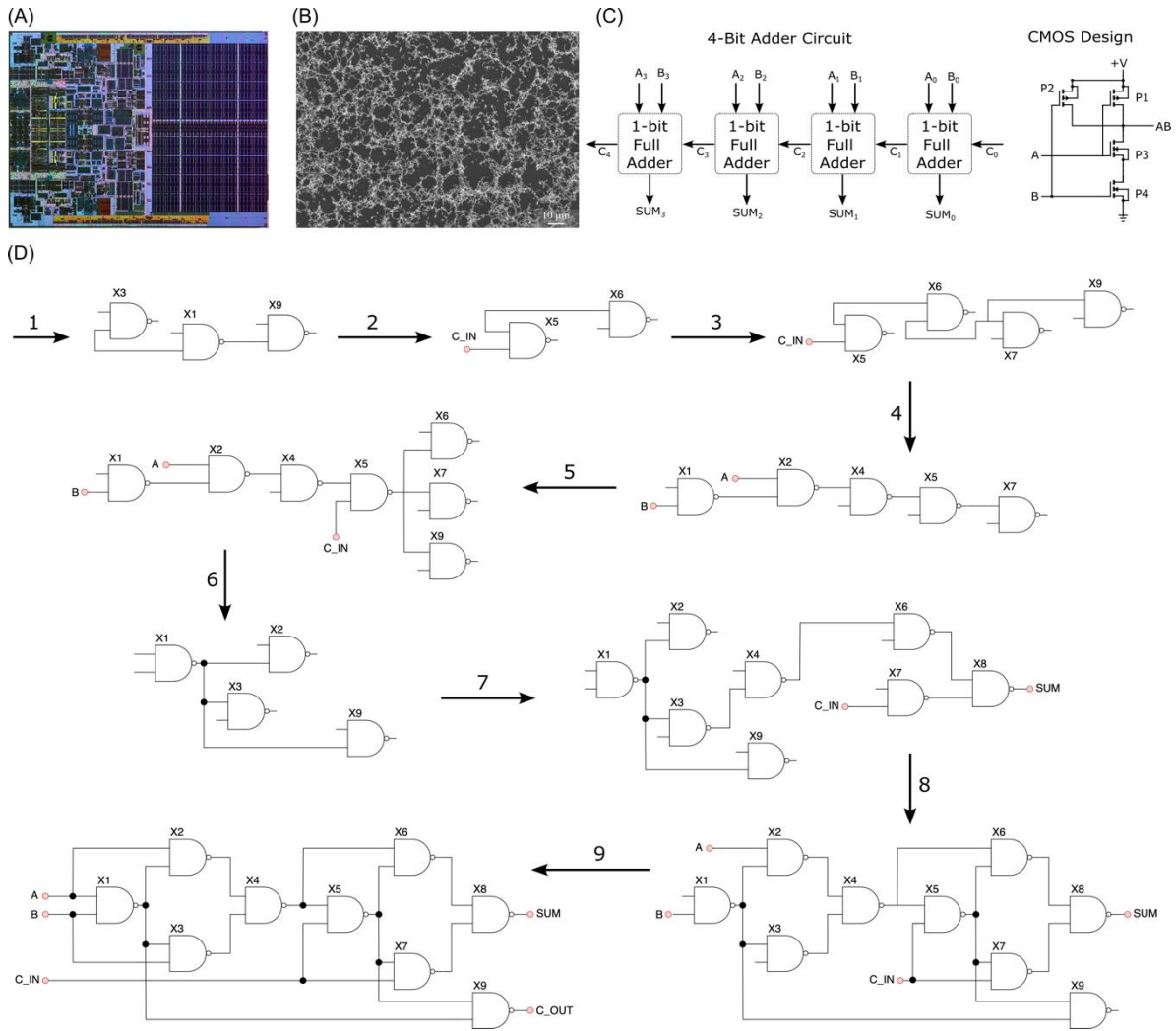

**Figure 1. Engineered vs. non-engineered architectures.** (A) shows the complex architecture of the integrated circuit of Intel's Xeon 3060 chip.[38] The well-engineered architecture shows a hierarchical construction that is not only periodic at atomic arrangements engineered with minimal defects but also well-designed at the nanometer and sub-micrometre scale. (B) SEM images of agglomerated silicon dioxide ($SiO_2$) nanoparticles under the influence of electromagnetic field. $SiO_2$ particles are periodic at the atomic scale, however, the arrangement of particles is highly disordered. (C) Schematic diagram of combining four 1-bit adders to create a 4-bit adder and its CMOS implementation. (D) Approximate assembly space of a 1-bit composed of nine NAND gates using a NAND gate and terminal (shown in red) as building blocks (see SI for detailed representation of pathway).

**Assembly Index of Periodic Crystalline Solids**

The principles of AT rely on the precise definition of an object. Building an assembly space is only justified when the construction of the object can be investigated using both theoretical and experimental techniques[39]. This means the assembly index only has meaning when it can be associated with some evidence of external control. For example, in the case of molecules, the governing conditions of the chemical reactions (reagents, process conditions, and catalysts)[40] control the formation of covalent bonds. In molecular assembly theory, it is straightforward to apply the concept of an object to isolated molecules, because these can only ever be detected when they are in high copy numbers. The assembly index relates to the formation of bonded interactions in molecular assembly theory. It can be used to quantify the amount of causation 'trapped' in each molecule, with the underlying assembly space capturing causal relationships connecting different molecules.

Here, we develop a new application of AT to objects which are composed of hierarchically connected unit cells (or irreducible units) in one-, two-, or three-dimensions such as nanoscale (or microscale and higher) particles, inorganic clusters and self-assembled or engineered structures. Most variations in solid-state objects considered here are present at atomic, or few nanometre scales such that larger-scale objects (microscale or higher) are inherently considered periodic. We assume that the assembly index of a large periodic object can be subdivided into constructing the minimal irreducible unit or asymmetric unit found within the crystallographic unit cell and combining these to form the observed material sample. Thus, for solid-state materials, the assembly space will constitute two nested parts: the shortest path to construct the unit cells with bonds as building blocks and the shortest path to construct the object with unit cells as building blocks. For simplicity, we define the assembly index of a crystalline object ($a_{obj}$) as a combination of two components: the assembly index of the unit cell with bonds as building blocks ($a_{uc}$) and the assembly index of the periodic arrangement of unit cells ($a_p$).

$$a_{obj} = a_{uc} + a_p \qquad (2)$$

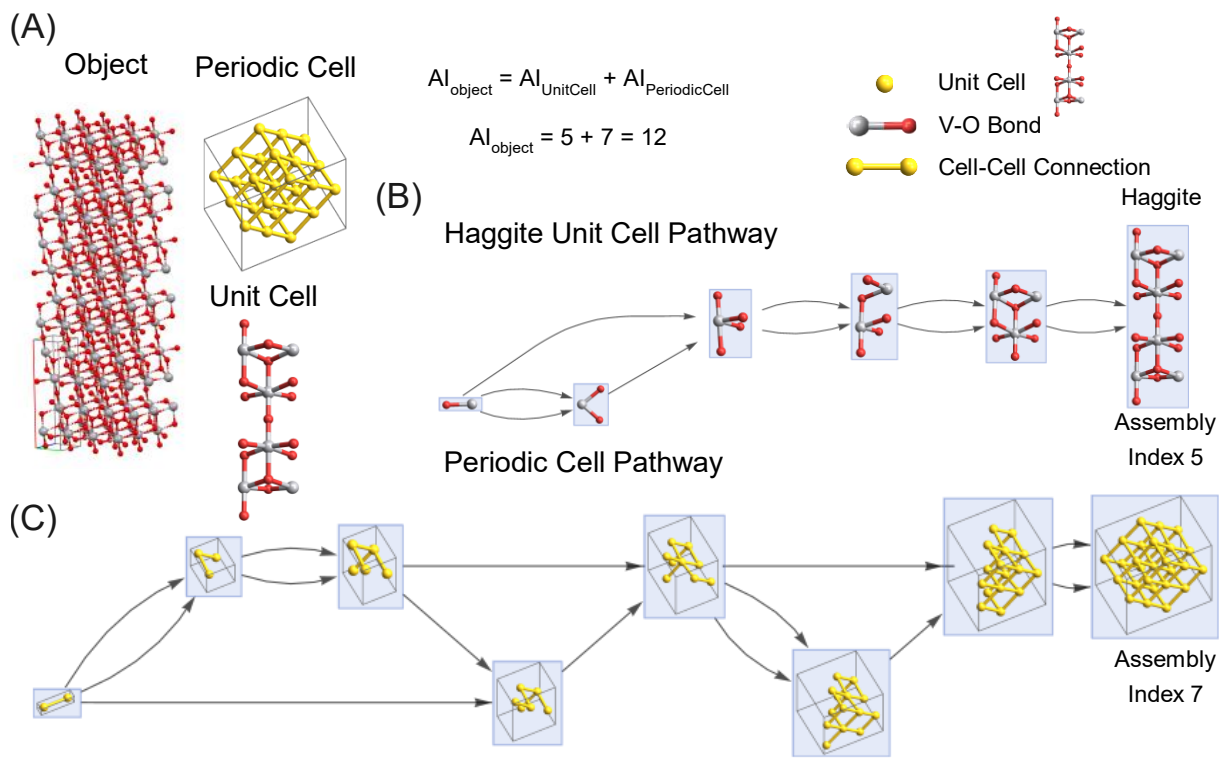

**Figure 2. Assembly Space of a periodic object.** (A) Figure shows a Haggite (3×3×3) crystal as an object. The object assembly is subdivided into two components: unit cell assembly and periodic assembly of unit cells. The overall assembly index is the summation of unit cell assembly index (5) and periodic cell assembly (7) and is equal to 12. (B) Assembly pathway to construct Haggite unit cell using V–O bond as a building block. (C) Assembly pathway of 3×3×3 periodic unit cell. (see SI Section 1 for details of calculating assembly indices).

This concept is shown in Figure 2A, which shows the division of assembly space into two components and estimation of the assembly index of a crystal of Haggite; here with a 3×3×3 unit with the *V–O bond* unit as a building block to estimate the assembly index of the unit cell (Figure 2B). The periodic cubic assembly of 3×3×3-unit cells is represented *by unit cell - unit cell* connection as a building block (Figure 2C). This hierarchical construction of the object from unit cells keeps the unit cell at an intermediate scale between the full assembled object, and bonds as its most elementary building blocks. It is not necessary for an object to have a single type of unit cell, as we show later a complex periodic object with defects can be represented by introducing supercells which are like unit cells but with larger volume. The nested space is necessary to define a causal connection between the object and the building blocks, as a unit cell or super cell is required to be constructed whose periodic arrangement then constitutes the object. So, in this regard, the intermediate object such as unit cell or super cell can be considered as an *internal object* and we define the number of copies required to constitute the object as the *internal copy number*, which is discussed in detail in the later section.

**Distribution of Assembly Indices of Known Minerals**

Previously, it was shown that molecular complexity can be used as a biosignature which can be experimentally measured.[29,41] A key result was the empirical demonstration via spectroscopy of a threshold in molecular assembly, above which the observation of complex molecules were only those created by an evolutionary process and uniquely associated to presence of life. While there has been many proposals of mineral-based biosignatures,[42] and even that minerals themselves might be considered evolutionary systems,[11] this has so far lacked empirically grounding to experimentally test these ideas. It remains an open question whether there is a threshold in complexity above which a mineral might be considered a biosignature or technosignature. Later we show theoretical motivation for a technosignature threshold that may be testable in the lab. Inspired by molecular assembly's use as a biosignature, we used the American Mineralogist Crystal Structure Database (AMCSD)[37] and calculated the assembly indices of the unique structures present in the database. The database of CIF files was filtered and, using CCSD Mercury Python API [43,44] and RDKit,[45] and then converted into mol files that represent the unit cells.

The complete pipeline to filter the database, including how to export the mol files of the unique crystal structures, is shown in detail in SI Section 2. The assembly index was then calculated on the exported mol files using our assembly calculator. Due to the high periodicity present in these structures, we developed a new assembly algorithm using a breadth-first approximation for the calculations written in C++, see SI Section 1 for details. The exact algorithm[46] enumerates all possible duplicatable subgraphs at the start of the calculation. This is not memory efficient in arrangements with a high degree of symmetry and scales exponentially with the number of bonds within the system. Instead, the breadth-first approximation algorithm attempts to find such subgraphs by performing a breadth-first search from each atom in the system. The breadth-first search is repeated on the list of subgraphs which possess duplicates until no subgraphs with duplicates remain. The structures found after this process are then recursively deleted from the parent graph in a manner similar to that we have described earlier[46]. This process is repeated until the breadth-first search fails to find any duplicatable structures greater in size than a single bond. Once the breath-first search is exhausted, the assembly index of the remaining graph fragments is calculated using the exact assembly algorithm described before[46] and the sum of the assembly index of both sub-algorithms is used to approximate the assembly index of the entire structure.

The distribution of assembly indices of the unit cells of various minerals from the database, using molecular bonds as building blocks, is shown in Figure 3A. In principle, the upper bound (shown in red) scales linearly with the number of bonds in a unit cell, which represent highly heterogeneous atomic arrangements such as high entropy alloys. Similarly, the lower bound (shown in red) scales logarithmically representing homogeneous and fully periodic unit cells in one-, two-, or three-dimensions. The distribution of assembly indices of mineral unit cells shows significant variations due to the number of bonds in the unit cells and their heterogeneity, with most unit cells tending to stay closer to the lower limit than the upper limit. Figure 3B shows a comparison between low and high-assembly index mineral unit cells from the database with a similar number of bonds within the unit cell. This analysis demonstrates strong variations in the heterogeneity of the unit cells leading to differences in complexity defined by their assembly index.

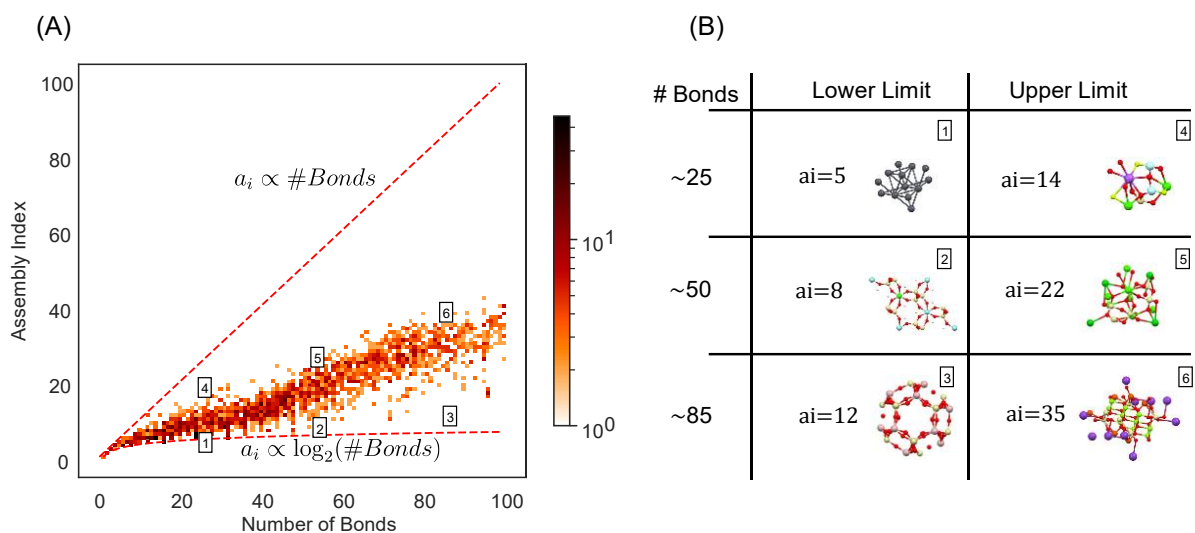

**Figure 3. Assembly index distribution of crystal structures** (A) Distribution of assembly index against the number of bonds, based on 4.3k structures sampled from the American Mineralogist Crystal Structure Database. The theoretical upper limit is linear, and lower limit logarithmic in the number of bonds. The assembly index values were calculated with the exact algorithm and the breadth-first search-based algorithm for low and high numbers of bonds, respectively. (B) Sample illustrating features of structures in different bond ranges. The characteristic features of a lower assembly index are very symmetric structures for a given number of bonds. The high assembly index usually comprises more asymmetric structures and a higher number of atom types.

Figure 3A shows a continuum of values, with no threshold in the data apparent, that might distinguish evolutionarily produced materials.[47,39] As we will show, this is because the copy number of structures within the material, inclusive of defects, is not yet accounted for, e.g., the size scale of the second term in Equation 2 will be determined by defects (See Section: Complexity of crystalline structures with point defects) in physical materials. Importantly, the assembly indexes in Figure 3A are theoretically determined, and it is unlikely that it will be possible to determine exact values for unit cells experimentally.[39] This is because, at the bonding level in solid-state materials, the crystallographic analysis averages over an entire crystal. This is contrasted with that of the microprocessor shown in Figure 1, where there are clear, countable, repeating structure within the material. In natural materials, defects and other disorders are averaged, meaning a precise assembly index and copy number do not have experimental meaning. This marks a large departure from the causation captured in the formation of complex molecules with a high assembly index and copy number, where exact values are experimentally probable. By contrast, mineral unit cells may have incredibly high assembly indices, but these also often only exist in the presence of high defect density. The causation captured by mineral fragments is very limited, and not directly probable, meaning new approaches are needed to quantify the complexity of crystalline materials.

**Joint Assembly Spaces Under Structural Transformation**

Phase transformations occur in the change of states of matter under different thermodynamic conditions such as temperature, pressure changes etc. An example from condensed matter is the eutectoid reaction in steel where a single solid phase ($\gamma$-Fe austenite) transforms into two solid phases ($\alpha$-Fe ferrite and $Fe_3C$ cementite). The composition of the two phases, ferrite and cementite are very different, and the phase transformation process starts via heterogeneous nucleation, followed by growth. The crystal structures of $\alpha$-Fe, $\gamma$-Fe, and $Fe_3C$ are body-centred cubic, face-centred cubic, and orthorhombic. The two-phased structure of Pearlite ($\alpha$-Fe 87.5 wt% and cementite 12.5 wt%) forms a lamellar structure as an outcome of the phase transformation. Using AT, complexity can be determined by considering the different unit cells, and the shape of the observed phases. This approach allows quantifying phase transformations as a contingent process defined by assembly space without the need of fully mechanistic process with high spatial and temporal information. By observing the Joint Assembly Space which represents the minimalistic pathway required to simultaneously construct an ensemble of objects, we can quantify the relationship between the observed phases during the phase transformation, in terms of the causation and constraints trapped within the material and how these transforms during the transition.

As an explicit example, we consider the phase transformation between two allotropes of carbon: graphite and diamond.[48] Graphite (or Graphene when considering a single layer) and diamond have different crystal structures, which are hexagonal and diamond cubic respectively, due to a different valence of the carbon atoms ($sp^2$ and $sp^3$) in the lattice. Hence, the structural transformation from graphite to diamond requires the transformation of carbon atoms from valence state $sp^2$ to $sp^3$. To quantify the transitions in constraints within the materials moving between phases of carbon allotropes, we created the joint assembly space of graphite, diamond, and three intermediate structures (gra_crbl33, gra_crbl43, diam_cr44), see Figure 4. We use C–C bonds as building blocks for the assembly space, using three different bonds as building blocks $C_{sp^2}-C_{sp^2}$, $C_{sp^2}-C_{sp^3}$, $C_{sp^3}-C_{sp^3}$. Here, we are only considering structural features involving different bond types on the unit cells and hence, the conservation of the total number of atoms during the structural transformation is not considered to construct the joint assembly space. The overlaps between the assembly spaces of graphite, diamond, and three intermediate structures within the joint assembly space signify the loss of contingent constraints within the material and the discovery of novel ones, during the phase transition where restructuring of the material leads to very different properties across the phase boundary, reflective of different sets of trapped constraints, as shown from left to right in Figure 4.

The transition from graphite to diamond shows a loss of contingency during the transition leading to the loss of causal structures involving $sp^2$–$sp^2$ bonds only after the first intermediate gra_crbl33. At this stage, during the formation of the second intermediate gra_crbl43, new bond types appear representing a discovery process within the assembly space, where the emerging assembly space consists of objects with three different bond types $sp^2$–$sp^2$, $sp^2$–$sp^3$, $sp^3$–$sp^3$. As the transition proceeds further towards diamond, the causality associated with $sp^2$–$sp^2$ and $sp^2$–$sp^3$ bond types is lost completely leading to new structures involving $sp^3$–$sp^3$ bonds only (diam_cr44 and diamond).

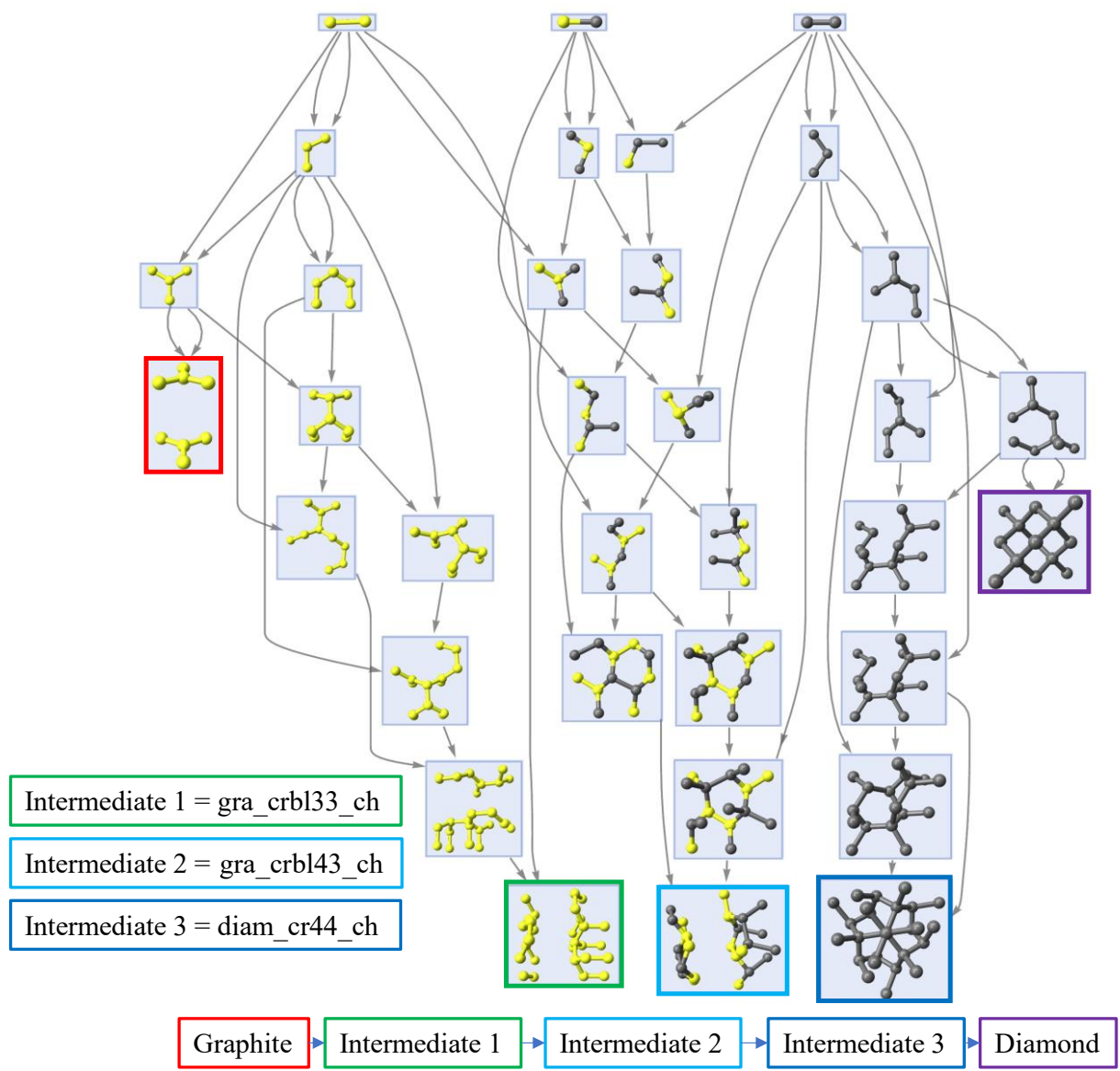

**Figure 4. Joint Assembly Space of Carbon Allotropes.** A proposed assembly tree and a joint assembly space between possible phases of graphite and diamond. The notation (cr) refers to a crystal structure, (bl) to a bilayer crystal and (ch) to the benzene from a graphite structure deforming to a chair-like arrangement in the intermediate structure. The joint assembly space captures the causal relationship from the structural features in the co-assembly space of all the structures in the phase transition of the different allotropes.

By comparing and quantifying the overlaps between the pathways of the observed objects within the JAS, physical mechanisms, like the phase transition of graphite to diamond, can be understood in new ways by highlighting the significance of trapped constraints within contingent pathways for their assembly, and how these transforms moving between phases.

**Periodic unit cells arrangement as an internal copy number**

AT utilises two observables, assembly index and copy number, to quantify the degree of selection in a physical system. In the case of crystalline materials, the boundary of an object is not clear, and the definition of a copy number is therefore not explicit. Copy number will depend on the resolution of the measurement as well as its associated error. The size of a solid-state object can be observed directly using high-resolution electron microscopy measurements (such as HR-TEM) or observing broadening in the powder X-ray diffraction experiments. The dependence of size broadening on the crystallite size is well-defined by the Scherrer equation.[49] As an example, consider spherical defect-free ZnS nanoparticles as objects in an ensemble, two nanoparticles of size 5nm and 6nm are highly similar and can be considered as countable copies of the same unique object type, as compared to two nanoparticles of size 5nm and 50nm, which are highly distinct. The periodicity in solid-state materials exists at two different scales, at the atomic scale (unit cells) and object scale (an entire sample), characterized by its size and shape. As mentioned previously, the unit cell defines the intermediate scale object(s), which connect bonds as building blocks with a shortest path to construct the unit cell, to then assemble the entire object with unit cells as the building block. That is, the material is treated as a nested hierarchy, and we introduce the concept of internal copy number to distinguish unit cell periodicity (unit cells repeating within a material sample) from object level periodicity (many samples of the same material). The internal copy number quantifies the number of unit cells required to construct the object. Additionally, using the intermediate scale objects, such as unit cells, to algorithmically construct the entire object contributes to the assembly index as given by equation 2, see Fig 2C for an example of an assembly space with an intermediate object that is a unit cell.

The periodicity within an object provides direct evidence of causal contingency, marking how once a unit cell is created, the information can be used recursively to construct the periodic object (conditions exist for repeated formation of the same unit cell). Considering a large periodic object (e.g., a naturally occurring mineral or engineered silicon chip), the assembly index $a$ can be defined as a summation of the assembly index of a single unit cell and the periodic arrangement of unit cells as given by equation 2. In general, for a cubic object, the number of steps ($s$) on the assembly pathway that combine unit cells to make the full object, scales logarithmically with the size of the object and its first-order approximation is given by $s \sim k \log_2(n)$, where $n$ is the number of repeated unit cells along an axis and $k$ is the dimensionality of the system. Hence, the assembly index can be described to scale roughly as $a \sim a_{uc} + k \log_2(n)$. At this point, it is important to note that as given by equation 1, for an object of finite size, the Assembly scales as $A \propto e^a$ where $a$ is the assembly index of the object. Considering a single object with a large periodic arrangement, the assembly $A$ is then approximated as $A \propto e^{a_{uc}} n^k$, where $n^k$ is the total number of unit cells in $k$ dimensions, i.e., $n^k$ is equivalent to internal copy number. This signifies that for a single extended periodic object, the object's Assembly $A$ is linearly proportional to internal copy number. Like equation 1, in cases where there is a single copy of an object, that is a unit cell, this would not represent contingency in the formation process (it is not repeatable), so we modify the Assembly accordingly to $A \propto p\, e^{a_{uc}} (n^k - 1)$.

The physical intuition comes from the fact that even considering the homogenous nucleation process, the critical radius ($r_c$) estimated from thermodynamic conditions for metals such as Cu is of the order of 1–3nm depending on the degree of supercooling. A stable nucleus larger than the critical radius is required for growth, including other factors such as temperature and supersaturation. Even the critical radius of homogenous nucleation is made of an array of unit cells (assuming $r_c = 1$nm and the copper lattice constant 0.361 nm, the number of unit cells is given by $n_{uc} = V_n/V_{uc} \approx 89$ where $V_n$ is the volume of the nucleus and $V_{uc}$ is the volume of the unit cell). A single unit cell does not persist long enough to have a causal effect on the growth dynamics of nucleation and therefore cannot be observed as an isolated object. Instead during nucleation, the critical size causal to growth comprises larger number of unit cells. Hence, $(n^k - 1)$ is justified for the quantification of Assembly, capturing that we expect the critical size of nucleation to never be an individual unit cell.

For a periodic object in two or three dimensions, the assembly space can be approximated using a tree data structure. Using a quadtree and octree decomposition in two and three dimensions, an object can be hierarchically constructed into larger super-objects recursively, starting from a single unit cell, which is implemented as the irreducible building block for a periodic arrangement of unit cells. The super-objects are therefore the elements with the most recursivity, constructed till the largest super-object fit within the actual object. An example of generating assembly spaces for circular (non-cubic) objects, using hierarchical decomposition, is shown in Figure 5. (see SI Section 2 for more details and examples).

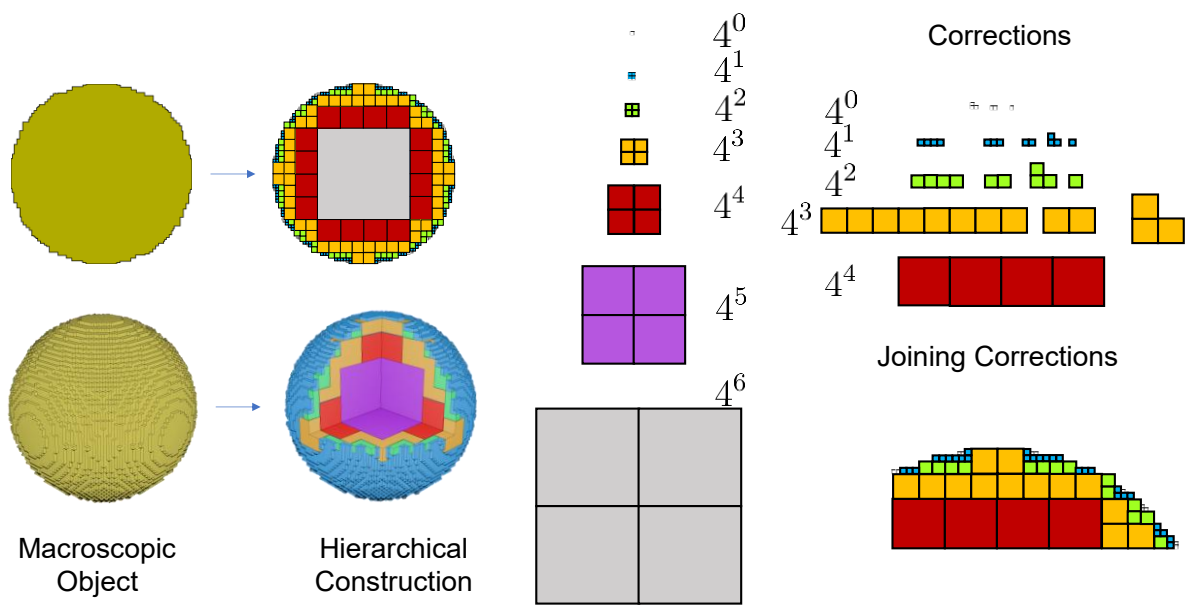

**Figure 5. Periodic arrangement of unit cells as internal copy number.** (A) shows the hierarchical decomposition of a discrete object recursively using quadtree and octree decomposition in two and three dimensions. (B) Approximate building blocks to create a circular object using quadtree decomposition[50] together with correction (additional objects) and additional joining operations.

**Assembly of Crystalline Structures with Point Defects**

We have considered idealized periodic structures that lack crystallographic defects. Compared to fully periodic lattices, quasicrystals, with their aperiodic tiling, are more 'complex' as they lack periodicity but are still ordered. Additionally, highly complex structures can be created from a periodic lattice by introducing crystallographic defects such as point defects (e.g. vacancies and interstitial defects), line defects (e.g. dislocations), and planar defects (e.g. grain boundaries and stacking faults). We next extend our analyses of crystalline materials using AT, by introducing point defects within crystalline lattices to account for internal copy number allowing us to quantify the complexity of real materials moving beyond other approaches focused solely on idealized examples of repeating unit cells.

Point defects such as vacancy or interstitial defects give rise to local disorder in the system, leading to higher complexity as compared to purely periodic lattices. The more the local disorder in the system, the more information is required to describe the lattice. AT can track the local disorder by considering different building blocks with and without defects and using these to determine the causation necessary to construct the observed object. Supercells, which are also periodic unit cells representing larger volumes, are often used to model point defects such that periodic boundary conditions (PBC) can be sustained. Here, we utilise the concept of supercells to quantify periodic lattices with local disorder defined by point defects. The assembly index of perfect, defect-free crystalline materials is calculated by summing the assembly index of a single unit cell and the assembly index of the periodic arrangement of the unit cells within the object (see equation 2). To introduce point defects characteristic of real materials, we consider two types of supercells with and without point defects, where the assembly index is defined as a summation of assembly indices of the two supercells and the arrangement of the two supercells to represent the entire crystalline material with defects, see Figure 6A.

We considered a synthetic crystalline lattice that can be generated by the periodic repetition of a cross as a unit cell $N \times N$ times and $N \times N \times N$ in 2D and 3D respectively. Next, we introduced point defects as vacancies in this lattice, where the resulting lattice can be analysed by an equivalent two-colour grid of $\frac{N}{2} \times \frac{N}{2}$ and $\frac{N}{2} \times \frac{N}{2} \times \frac{N}{2}$ in 2D and 3D, respectively. These two colours represent the two types of supercells that generate this lattice. For large $N$, assembly index calculations using the breadth-first search-based algorithm are time consuming and memory-intensive. Instead, we approximate the assembly index of this square and cubical grid using the Hash-Assembly algorithm[51]. This algorithm generates a quadtree or octree decomposition of the grid and generates an approximate assembly space from the hashing of the nodes of the previous trees, see SI Section 3 for more details and SI Section 1 for a comparison with other algorithms. This approximation is shown to be accurate to the first order for our application since it was adapted from the Hash-Life algorithm which is designed to take advantage of the regularities of large sparse cellular spaces[52]. We estimate the assembly index using a Monte-Carlo estimate of 60 runs with $\frac{N}{2}$ equal to 16, 64 and 512 and a sweep of point defect densities, this is shown in Figure 6b. We observe a rise in assembly index with the increase in point defects, and there is also an effect from the dimensionality of the system that constrains the locations of the point defects given the imposition of no voids (no two-point defects lie exactly next to each other).

**Assembly Spaces of Hexagonal Closed-Packed Structures with Stacking Faults**

Planar defects, such as stacking faults, occur in crystalline layered materials during crystal growth and are called growth faults, or can occur in plastic deformation and are called deformation faults. These faults are often randomly distributed over the layered structure. Stacking faults also occur during phase transformations between closed packed structures such as 2*H* to 6*H* and are known to be non-random. Analysis of these phase transformations between hexagonally closed-packed (HCP) and face-centred cubic (FCC) structures have explored the effect of fault density on the observed diffraction intensity.[53] We next use AT to quantify the complexity of these emerging layered structures, based on observable fault probabilities. Additionally, we demonstrate how Assembly can differentiate systems where faults are generated by a random or *engineered* process. To do so, we perform Monte Carlo simulations of random growth of stacking faults and quantify their properties using AT[32].

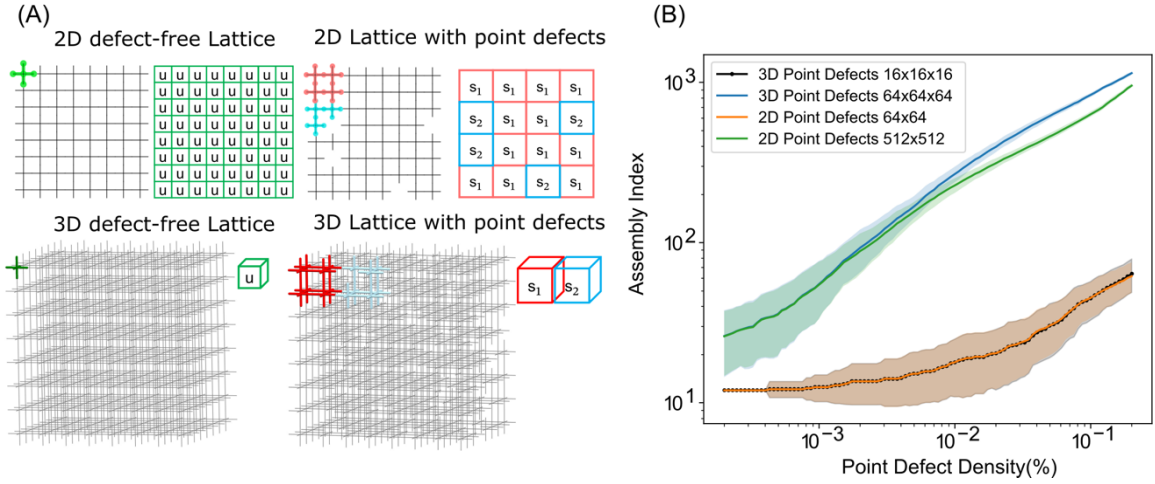

**Figure 6. Supercell approximation of a crystal with point defects.** (A) Periodic structure in 2D and 3D from a simple unit cell and introduction of point defects in the structures in the form of vacancies. We introduce a supercell approach, mapping the problem of the assembly index of an asymmetric graph to the assembly index of a 3D box with black and white squares. (B) Variation of the assembly index of different sized 2D or 3D boxes given the point defect density, the assembly index increases as the object becomes more asymmetric.

Like the last section, here we consider a single observable object and define an intermediate length scale to define the internal objects and their copy numbers. Consider an ensemble of layers arranged in an HCP structure of $10^4$ layers, where we consider $L = 2^8 \times 39 = 9984$ such that it is easily represented by a binary tree. The fault-free HCP and FCC layers are defined as $ABABABAB$ ... and $ABCABCABCABC$ ... respectively. A single stacking fault can be introduced using an anti-cyclic transformation rule: $A \to C, B \to A, C \to B$ on the HCP lattice. As an example, a single fault introduced in the HCP layer $ABABAB|ABAB$ ... will lead to $ABABABCACA$ ... where $ABAB$ ... and $CACA$ ... represents the HCP phase and $ABC$ represents phase transformation to FCC phase.

To explore stacking faults we started with perfect HCP structures defined by an interval [1,9984] by setting the fault probability $\beta = 0.25$, yielding 2496 faults. The approach considers that sites that have been used previously are blocked, and no two successive sites comprise the same layer type. The pair correlation functions $P(m)$, $Q(m)$, and $R(m)$ represents the probabilities of observing $A$–$A$, $B$–$B$, $C$–$C$; $A$–$B$, $B$–$C$, $C$–$A$; and $A$–$C$, $B$–$A$, $C$–$B$ pairs at a separation of $m$ layers. It has been shown previously[53] that with an increase in the separation, the probabilities fall exponentially, with a functional form $P(m) = a \mp b\, e^{-m/l}$ where $l$ is the characteristic length scale that is itself a function of the fault probability $\beta$. Using the pair correlation functions, we define an internal object length scale ($\xi$) within the material, and perform assembly index calculations, such that assembly indices estimates are computationally tractable and the internal copy number is quantifiable. For a closed-packed layer with a given fault probability, the internal object length scale is defined as the length scale such that the correlation probability is less than 1%. This pair correlation analysis allows us to discern the internal object length scale, which then defines the internal object and copy number. Unlike molecules[32], which have well-defined boundaries, crystalline structures possess blurry boundaries, allowing objects generated with different dynamics to lie adjacent to each other. This approach is particularly useful when considering the size of crystalline materials, such that crystals are defined by lengths $L$ and $L + \delta$ with $\delta \ll L$ and these can be easily compared by considering their subparts defined by the internal object correlation length $\xi$. Hence, using the characteristic internal object length scale, we divide the given layered structure of length $L$ (number of layers in arbitrary length units) in sub-parts of length $\xi$, i.e. the internal correlation length, which defines the internal objects and their copy number. This allows obtaining a total of $N_T$ internal objects within the material. Using this approach, all key features of an object as defined in AT[32] are fulfilled, including distinguishability, finiteness, composability, and constraints allowing meaningful application of AT, not just assembly index, to quantify causation within materials.

We consider a closed-packed layered object (length $L \sim 10^4$) with five different fault probabilities, where each can represent internal objects of sizes given by their correlation lengths. Next, we generated an ensemble of internal objects for each size from the object and grouped the unique internal objects by their copy numbers and assembly indices to compute the material's Assembly. This process was repeated 60 times, and averaging the features, the observed distributions are shown in Figure 7. As a next step, we computed the Assembly for crystals with a parametric sweep of fault densities (which led to different correlation lengths for defining internal objects) and this was averaged over $10^5$ Monte Carlo simulations to obtain meaningful statistics for the material (See SI Section 4 for more details). At five different correlation lengths, we observe that for a crystal with fixed total length $L$, materials with smaller internal object correlation lengths ($\xi$) have a larger number of unique internal objects, which are in general less complex. Conversely, materials with large internal object correlation lengths have a smaller number of unique objects, but these tend to be more complex with higher assembly indices. This balance between the number of unique internal objects within in hierarchical material of length $L$, their assembly indices, and copy numbers represents fundamental trade-offs in how material complexity can be realized.

Assembly quantifies the degree of selection within the space of material constraints, which is required for a given material to be observed, and in this case represents the hierarchical object. Here, we explore the relationship of Assembly of materials with respect to the fault probabilities. We find Assembly has a region of maximum values at low fault densities, and for larger values, it decreases. If we zoom in on this region of maximum values, we observe a region where the Assembly oscillates around an approximate value of $10^8$; this region corresponds to (on average) the maximum Assembly of a crystal with random faults. Here, the region contains characteristic lengths that oscillate between $\xi \in [1.09\%L, 33.5\%L]$, see Figure 7.

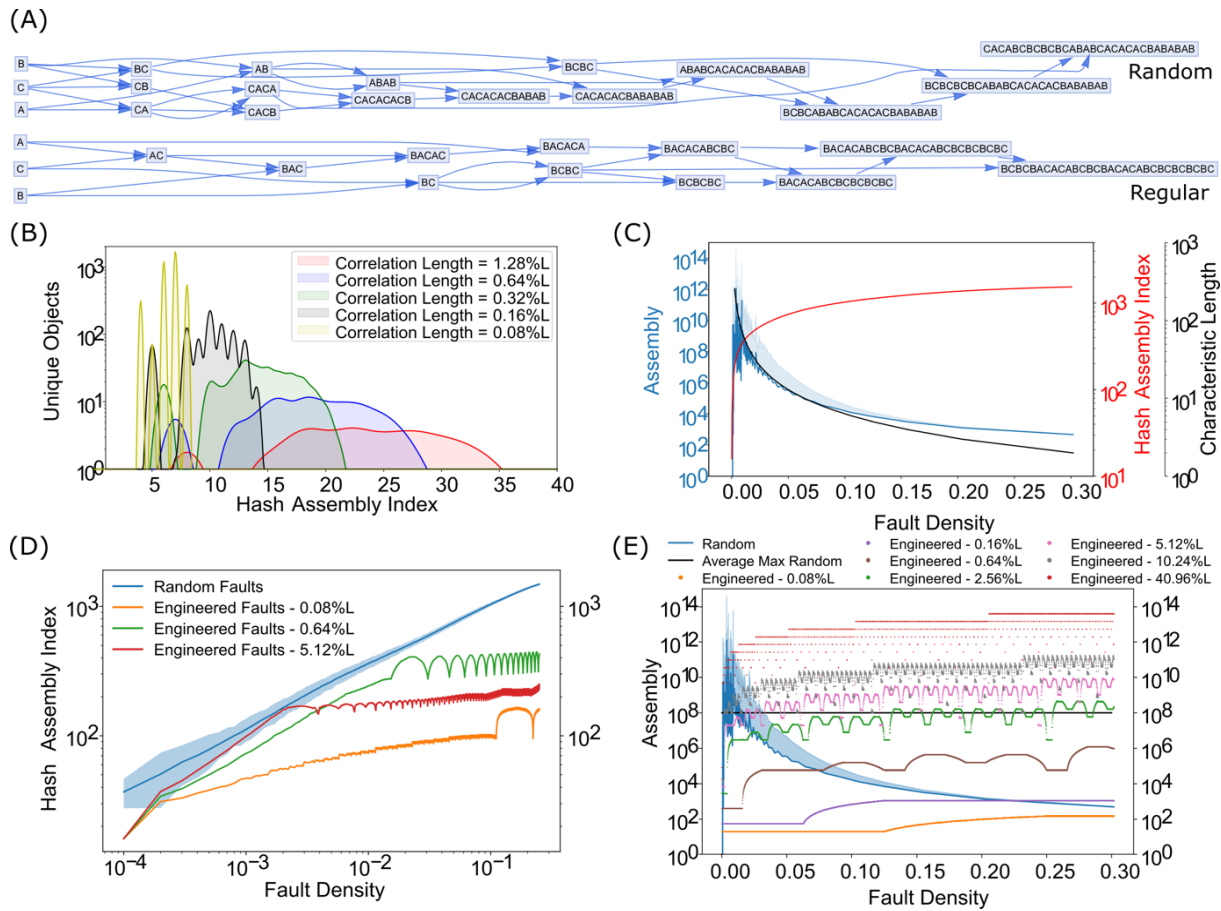

**Figure 7. Random and Engineered Stacking Faults in a Periodic Crystal.** (A) Assembly Space of a crystal of length of $L = 30$ with five faults in the random and periodic process. (B) Trade-off between the number of unique objects and their assembly indices in a crystal with layers $L = 10\,000$, showing those with a higher density of faults, or conversely with a longer correlation length. (C) Assembly of a crystal with a layer $L \sim 10\,000$ with respect to the density of faults. The Assembly for small fault density oscillates at an average maximum of $10^8$. (D) Assembly index of a crystal with layers $L = 10\,000$ undergoing different regular fault lengths, or random faults given a specific fault density. (E) Assembly of a crystal with layers $L = 10\,000$ undergoing different regular fault lengths, or random faults given a specific fault density. The engineered faults can surpass the average maximum value of $10^8$ given by the random faulted crystal, providing formalization of a material technosignature.

Previously, we have considered the stacking faults formed in closed-packed layers as an outcome of random processes. For a comparative study, we consider programmed faults at well-defined sites, where the observed crystal is *engineered*. We consider a crystal undergoing regular faulting. We introduce one stacking fault for every $\hat{\xi}$ number of layers. This generates a layered stack of regularly sized crystal structures, very much like the well-defined architectures found in commercial semiconductor devices[54]. We assume that this regular faulting process could be achieved with advanced manufacturing techniques[54,55] and therefore is characteristic of technological processes. As with the previous case, we consider an ensemble of layers arranged in a defect-free HCP lattice of length $L = 9984$.

By applying a similar transformation (anti-cyclic) as explained above, introducing one fault in every $\hat{\xi}$ number of layers until a fault density of $\beta = 0.25$ is achieved, e.g., 2496 faults in this example, where we have excluded sites that have already been used. We investigate the bulk properties of the crystal for three different regular faulting lengths $\hat{\xi} \in \{0.08\%L, 0.64\%L, 5.12\%L\}$. After computing the hash-assembly index of each ensemble of layers, for each regular length, and these were compared with the Monte Carlo simulation of a crystal with random faults, see Figure 7. For crystals with random faults, we found the hash-assembly index is always larger than the engineered crystals, and at this microscopic length scale, the random crystal is more complex than the engineered ones. However, macroscopic complexity is only observed in engineered materials, as reflected by their Assembly. Compression measures could likewise pick up complexity at the microscale, but lacking a physical, solid-state interpretation could not explain the difference in macroscale complexity for the engineered material. A general feature we observed was how the assembly index grows at each length until the crystal has a density of faults exactly equal to $\hat{\xi}/L$. The assembly index has a periodic behaviour with a bias toward higher values. However, it is not necessarily the largest regular lengths that acquire the largest hash-assembly indices, and this is something that necessitates further explanation.

Next we computed the Assembly values using equation (1) for crystals with correlation lengths $\hat{\xi} \in \{2^i \cdot 0.08\%L\}_{i=0}^{6}$ as well as for each fault probability up to a fault probability $0.3L$, and we compared these to the example with random faults. In doing this we find that engineered crystals can be more "complex" than a random crystal. This separation has a well-defined physical meaning: constraints from the dynamics of a random defect formation are not able to generate an internal ensemble of objects within a material that would trap enough causal contingency to generate macroscale order, a key differentiator of engineered crystals designed to store and process information. Thus, the engineering process introduces a strong selectivity that produces many copies of internal objects, where these internal objects have considerable complexity. The exponential dependency of Assembly on the assembly index in equation (1) remains the dominant term, even though an ensemble of objects with lower correlation lengths will have higher copy numbers of internal objects. Therefore, we conclude that a crystal with engineered faults with a high enough number of complex internal objects can reveal the selectivity inherent in its engineered design, and this distinguishes these materials *physically* (in a manner that is measureable) from a naturally formed crystal with random faults.

**Conclusions**

Here we showed how it is possible to extend the foundational principles of Assembly Theory (AT) to inorganic molecules, periodic crystalline structures, and engineered solid-state materials. To do this we advanced the theoretical framework of AT to develop a robust approach for quantifying the complexity and associated contingency in crystalline materials. This is important as we were able to develop a methodology to quantify assembly indices over large periodic structures by utilising a hierarchical approach. This meant that, by decomposing crystalline structures into unit cells and analysing their periodic arrangements with defects, it was possible to use AT to explore the complexity encapsulated from both local atomic-level order and large-scale structural periodicity. Indeed, by calculating the assembly indices of the unit cells of structures from large mineral databases, it was possible to demonstrate how the heterogeneity in bonding and atomic arrangement gives highly complex unit cells. Importantly, this does not necessarily mean the material itself is complex, with causal constraints manifest at the macroscale, as such a material would be described as macroscopically complex. However, we introduced the concept of internal copy number to differentiate between periodicity at the unit-cell level and within the entire material sample, which comprises the large-scale arrangement of unit cells. This approach allows us to efficiently characterize hierarchically organized materials. In so doing we can differentiate microscale complexity that can be driven from random processes, from macroscale complexity that arise through selection. This allowed us to demonstrated the potential of AT to discern material technosignatures as the product of engineering design, providing a quantitative basis for distinguishing natural processes from both evolutionary and intelligent design. Finally, our approach can be utilised to explore the interplay between complexity and functionality towards novel materials design. AT offers a systematic approach to navigate this balance, enabling the optimisation of materials for mechanical, electrical, and thermal properties. Future research will focus on extending the framework to encompass dynamic processes, such as real-time phase transitions and self-assembly in condensed and soft matter.

**Methods**

All the assembly index calculations using breadth-first approximations were performed using C++ and hash-assembly calculations were performed using Python 3. The data analysis was performed using Python 3 and Mathematica 14. Complete details of all the methods are described in detail in the Supplementary Information.

**Code Availability**

All the codes required for assembly calculations and generating the figures are available at https://github.com/croningp/solidstate_complexity.

**Author Contributions**

L.C. conceived the idea together with A.S. L.C. S.W., A.S. and K.Y.P developed the research plan and K.Y.P did the crystallographic analysis, phase transition analysis, and Monte-Carlo investigations. A.S. developed the application of pairwise corelation functions and worked with K.Y.P to develop the idea. I.S. developed the assembly calculator using breadth-first algorithm and I.P. did the initial investigations. K.Y.P, A.S., S.W. and L.C. wrote the manuscript.

**Acknowledgements**

We acknowledge financial support from the John Templeton Foundation (grant nos. 61184 and 62231), the Engineering and Physical Sciences Research Council (EPSRC) (grant nos. EP/L023652/1, EP/R01308X/1, EP/S019472/1 and EP/P00153X/1), the Breakthrough Prize Foundation and NASA (Agnostic Biosignatures award no. 80NSSC18K1140), MINECO (project CTQ2017-87392-P) and the European Research Council (ERC) (project 670467 SMART-POM).

**References**

1. Lehn, J. Perspectives in Chemistry—Steps towards Complex Matter. *Angew. Chem. Int. Ed.* **52**, 2836–2850 (2013).

2. Johnston, I. G. *et al.* Symmetry and simplicity spontaneously emerge from the algorithmic nature of evolution. *Proc. Natl. Acad. Sci.* **119**, e2113883119 (2022).

3. England, J. L. Dissipative adaptation in driven self-assembly. *Nat. Nanotechnol.* **10**, 919–923 (2015).

4. Saha, T. & Galic, M. Self-organization across scales: from molecules to organisms. *Philos. Trans. R. Soc. B Biol. Sci.* **373**, 20170113 (2018).

5. Anderson, P. W. More Is Different. *Science* **177**, 393–396 (1972).

6. Herman, M. A. *et al.* A Unifying Framework for Understanding Biological Structures and Functions Across Levels of Biological Organization. *Integr. Comp. Biol.* **61**, 2038–2047 (2021).

7. Eigen, M. & Schuster, P. Stages of emerging life —Five principles of early organization. *J. Mol. Evol.* **19**, 47–61 (1982).

8. Böttcher, T. An Additive Definition of Molecular Complexity. *J. Chem. Inf. Model.* **56**, 462–470 (2016).

9. Randić, M. & Plavšić, D. Characterization of molecular complexity. *Int. J. Quantum Chem.* **91**, 20–31 (2003).

10. Warr, W. A. Combinatorial Chemistry and Molecular Diversity. An Overview. *J. Chem. Inf. Comput. Sci.* **37**, 134–140 (1997).

11. Phillips, B. L., Casey, W. H. & Karlsson, M. Bonding and reactivity at oxide mineral surfaces from model aqueous complexes. *Nature* **404**, 379–382 (2000).

12. Galarneau, A., Barodawalla, A. & Pinnavaia, T. J. Porous clay heterostructures formed by gallery-templated synthesis. *Nature* **374**, 529–531 (1995).

13. Cairns-Smith, A. G. Chemistry and the Missing Era of Evolution. *Chem. – Eur. J.* **14**, 3830–3839 (2008).

14. Crick, F. H. C. The origin of the genetic code. *J. Mol. Biol.* **38**, 367–379 (1968).

15. Kauffman, S. A. & Roli, A. A third transition in science? *Interface Focus* **13**, 20220063 (2023).

16. Zhang, J. *et al.* Experimental constraints on the phase diagram of titanium metal. *J. Phys. Chem. Solids* **69**, 2559–2563 (2008).

17. Krivovichev, S. V. Which Inorganic Structures are the Most Complex? *Angew. Chem. Int. Ed.* **53**, 654–661 (2014).

18. Cronin, L. Exploring the Hidden Constraints that Control the Self-Assembly of Nanomolecular Inorganic Clusters. *Bull. Jpn. Soc. Coord. Chem.* **78**, 11–17 (2021).

19. Miras, H. N., Wilson, E. F. & Cronin, L. Unravelling the complexities of inorganic and supramolecular self-assembly in solution with electrospray and cryospray mass spectrometry. *Chem. Commun.* 1297–1311 (2009) doi:10.1039/B819534J.

20. Kaußler, C. & Kieslich, G. *crystIT* : complexity and configurational entropy of crystal structures via information theory. *J. Appl. Crystallogr.* **54**, 306–316 (2021).

21. Fang, Z. *et al.* Structural Complexity in Metal–Organic Frameworks: Simultaneous Modification of Open Metal Sites and Hierarchical Porosity by Systematic Doping with Defective Linkers. *J. Am. Chem. Soc.* **136**, 9627–9636 (2014).

22. Wei, L. *et al.* Encoding ordered structural complexity to covalent organic frameworks. *Nat. Commun.* **15**, 2411 (2024).

23. Hornfeck, W. On an extension of Krivovichev's complexity measures. *Acta Crystallogr. Sect. Found. Adv.* **76**, 534–548 (2020).

24. Krivovichev, S. V. Structural complexity and configurational entropy of crystals. *Acta Crystallogr. Sect. B Struct. Sci. Cryst. Eng. Mater.* **72**, 274–276 (2016).

25. Shannon Entropy - an overview | ScienceDirect Topics. https://www.sciencedirect.com/topics/engineering/shannon-entropy.

26. Kraskov, A., Stögbauer, H. & Grassberger, P. Estimating mutual information. *Phys. Rev. E* **69**, 066138 (2004).

27. Kolmogorov, A. N. Three approaches to the quantitative definition of information *. *Int. J. Comput. Math.* **2**, 157–168 (1968).

28. Kauffman, S. A. Origins of Order in Evolution: Self-Organization and Selection. in *Understanding Origins* (eds. Varela, F. J. & Dupuy, J.-P.) 153–181 (Springer Netherlands, Dordrecht, 1992). doi:10.1007/978-94-015-8054-0_8.

29. Marshall, S. M. *et al.* Identifying molecules as biosignatures with assembly theory and mass spectrometry. *Nat. Commun.* **12**, 3033 (2021).

30. Jirasek, M. *et al.* Investigating and Quantifying Molecular Complexity Using Assembly Theory and Spectroscopy. *ACS Cent. Sci.* **10**, 1054–1064 (2024).

31. Slocombe, L. & Walker, S. I. Measuring Molecular Complexity. *ACS Cent. Sci.* **10**, 949–952 (2024).

32. Sharma, A. *et al.* Assembly theory explains and quantifies selection and evolution. *Nature* **622**, 321–328 (2023).

33. Krivovichev, S. V. Which Inorganic Structures are the Most Complex? *Angew. Chem. Int. Ed.* **53**, 654–661 (2014).

34. Liu, Y. *et al.* Exploring and mapping chemical space with molecular assembly trees. *Sci. Adv.* **7**, eabj2465 (2021).

35. Chhugani, J. *et al.* Efficient implementation of sorting on multi-core SIMD CPU architecture. *Proc VLDB Endow* **1**, 1313–1324 (2008).

36. Brinkman, W. F., Haggan, D. E. & Troutman, W. W. A history of the invention of the transistor and where it will lead us. *IEEE J. Solid-State Circuits* **32**, 1858–1865 (1997).

37. Robert T. Downs & Michelle Hall-Wallace. American mineralogist crystal structure database. *Choice Rev. Online* **41**, 41Sup-0262-41Sup – 0262 (2004).

38. Central processing unit. *Wikipedia* (2024).

39. Walker, S. I., Mathis, C., Marshall, S. & Cronin, L. Experimentally measured assembly indices are required to determine the threshold for life. *J. R. Soc. Interface* **21**, 20240367 (2024).

40. Cronin, L. The Chemputer and Chemputation: A Universal Chemical Compound Synthesis Machine. Preprint at https://doi.org/10.48550/arXiv.2408.09171 (2024).

41. Jirasek, M. *et al.* Investigating and Quantifying Molecular Complexity Using Assembly Theory and Spectroscopy. *ACS Cent. Sci.* **10**, 1054–1064 (2024).

42. Banfield, J. F., Moreau, J. W., Chan, C. S., Welch, S. A. & Little, B. Mineralogical Biosignatures and the Search for Life on Mars. *Astrobiology* **1**, 447–465 (2001).

43. Allen, F. H. *et al.* The Cambridge Crystallographic Data Centre: computer-based search, retrieval, analysis and display of information. *Acta Crystallogr. B* **35**, 2331–2339 (1979).

44. Macrae, C. F. *et al.* Mercury: visualization and analysis of crystal structures. *J. Appl. Crystallogr.* **39**, 453–457 (2006).

45. Bento, A. P. *et al.* An open source chemical structure curation pipeline using RDKit. *J. Cheminformatics* **12**, 51 (2020).

46. Seet, I., Patarroyo, K. Y., Siebert, G., Walker, S. I. & Cronin, L. Rapid Computation of the Assembly Index of Molecular Graphs. Preprint at http://arxiv.org/abs/2410.09100 (2024).

47. Hazen, R. M. *et al.* Molecular assembly indices of mineral heteropolyanions: some abiotic molecules are as complex as large biomolecules. *J. R. Soc. Interface* **21**, 20230632 (2024).

48. Fayos, J. Possible 3D Carbon Structures as Progressive Intermediates in Graphite to Diamond Phase Transition. *J. Solid State Chem.* **148**, 278–285 (1999).

49. Holder, C. F. & Schaak, R. E. Tutorial on Powder X-ray Diffraction for Characterizing Nanoscale Materials. *ACS Nano* **13**, 7359–7365 (2019).

50. Shusterman, E. & Feder, M. Image compression via improved quadtree decomposition algorithms. *IEEE Trans. Image Process.* **3**, 207–215 (1994).

51. Patarroyo, K. Y., Sharma, A., Walker, S. I. & Cronin, L. AssemblyCA: A Benchmark of Open-Endedness for Discrete Cellular Automata. in (New Orleans, 2023).

52. Gosper, R. Wm. Exploiting regularities in large cellular spaces. *Phys. Nonlinear Phenom.* **10**, 75–80 (1984).

53. Tiwary, P. & Pandey, D. Scaling behaviour of pair correlation functions for randomly faulted hexagonal close-packed structures. *Acta Crystallogr. A* **63**, 289–296 (2007).

54. Campbell, S. A. *Fabrication Engineering at the Micro- and Nanoscale*. (Oxford Univ. Press, New York, 2013).

55. Rozgonyi, G. Silicon: Characterization by Etching. in *Encyclopedia of Materials: Science and Technology* 8524–8532 (Elsevier, 2001). doi:10.1016/B0-08-043152-6/01522-9.